\documentclass[preprint]{vgtc}               
\ifpdf
  \pdfoutput=1\relax                   
  \usepackage{graphicx}                
  \DeclareGraphicsExtensions{.pdf,.png,.jpg,.jpeg} 
\else
  \usepackage{graphicx}                
  \DeclareGraphicsExtensions{.eps}     
\fi%

\graphicspath{{figures/}{pictures/}{images/}{./}} 

\usepackage{microtype}                 
\PassOptionsToPackage{warn}{textcomp}  
\usepackage{textcomp}                  
\usepackage{mathptmx}                  
\usepackage{times}                     
\usepackage{cite}                      
\usepackage{tabu}                      
\usepackage{booktabs}                  

\usepackage{enumitem}
\usepackage{setspace}

\usepackage{url}

\vgtcinsertpkg

\title{CleanAirNowKC: Building Community Power by Improving Data Accessibility}

\author{Rifat Ara Proma
\thanks{e-mail: proma@usf.edu}\\ %
     \scriptsize University of South Florida %
\and Matthew Sumpter
\thanks{e-mail: mjsumpter@usf.edu}\\ %
     \scriptsize University of South Florida %
\and Humberto Lugo
\thanks{e-mail: betomtz.lugo@gmail.com}\\ %
     \scriptsize CleanAirNow %
\and Elizabeth Friedman 
\thanks{e-mail: efriedman@cmh.edu}\\ 
     \scriptsize Children’s Mercy, Kansas City %
\and Khandaker Tasnim Huq
\thanks{e-mail: kthuq@usf.edu}\\ %
     \scriptsize University of South Florida %
\and Paul Rosen
\thanks{e-mail: prosen@usf.edu}\\ %
     \scriptsize University of South Florida %
}

\teaser{
  \centering
  {\includegraphics[trim=0 10pt 10pt 57pt, clip, width=0.95\linewidth]{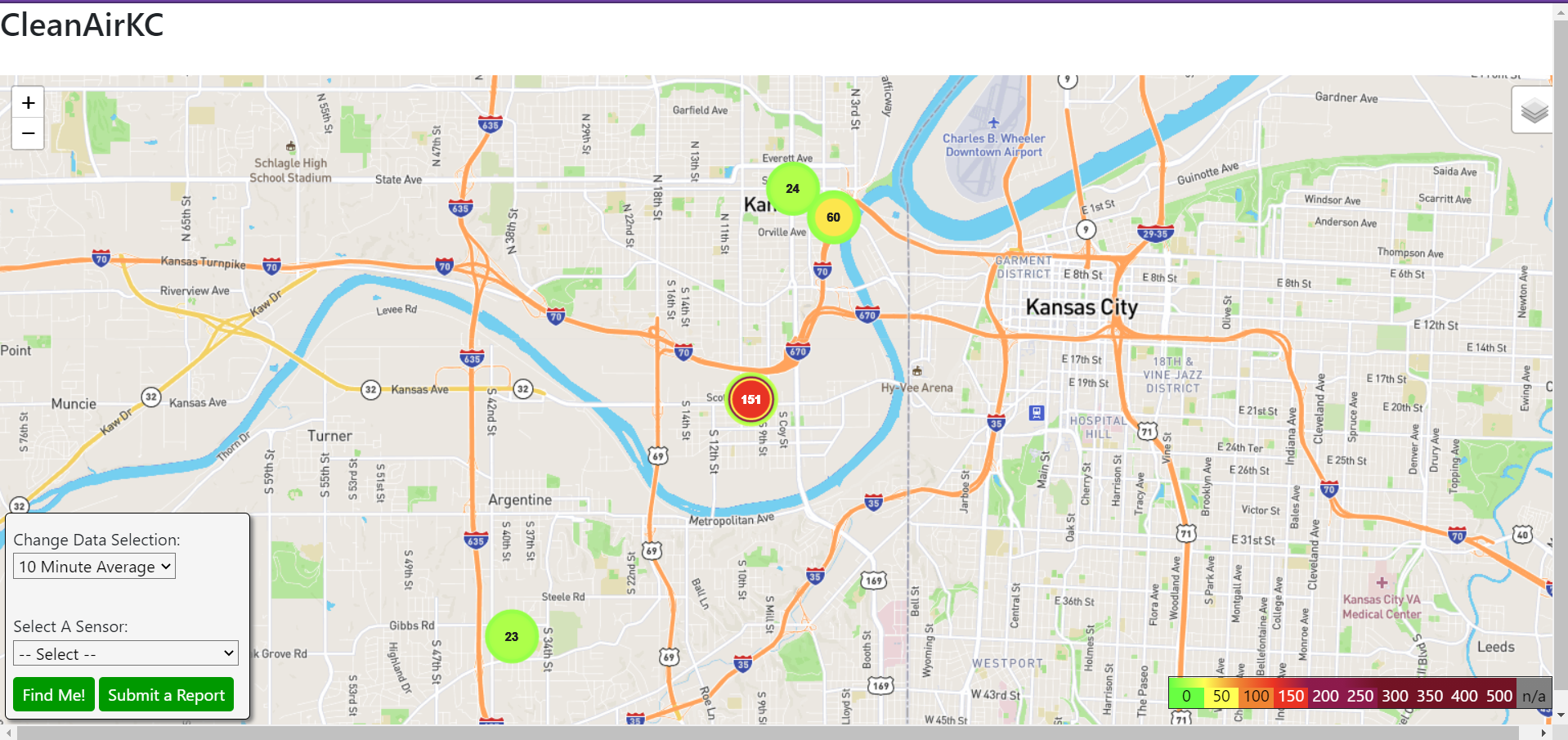}}
  \caption{CleanAirNowKC full map displayed on loading the website.}
  \label{fig:map}
}

\abstract{As cities continue to grow globally, air pollution is increasing at an alarming rate, causing a significant negative impact on public health. One way to affect the negative impact is to regulate the producers of such pollution through policy implementation and enforcement. CleanAirNowKC (CAN-KC) is an environmental justice organization based in Kansas City (KC), Kansas. As part of their organizational objectives, they have to date deployed nine PurpleAir air quality sensors in different locations about which the community has expressed concern. In this paper, we have implemented an interactive map that can help the community members to monitor air quality efficiently. The system also allows for reporting and tracking industrial emissions or toxic releases, which will further help identify major contributors to pollution. These resources can serve an important role as evidence that will assist in advocating for community-driven just policies to improve the air quality regulation in Kansas City.%
} 

\begin{document}

\firstsection{Introduction}

\maketitle

Rapid urbanization is occurring globally, and with it comes an increased threat of air pollution~\cite{parrish2015urbanization}. Urbanization leads to more centralized industries, including transportation, waste management, and energy production~\cite{Landrigan2017}. Air pollution has been shown to have significant adverse effects on public health, including an increased risk of asthma~\cite{martenies2018effectiveness}, cardiovascular disease~\cite{commodore2017community}, and cancer~\cite{turner2017ambient}, leading to an estimated healthcare cost of US\$21 billion globally in 2015~\cite{economicpollution}. Air pollution is an avoidable consequence of urbanization; however, in order to regulate the main producers of pollution, significant policy implementations and enforcement are required~\cite{Landrigan2017}. 

Air pollution policy can only succeed with a high level of community engagement, and those most likely to become engaged are those who are most directly impacted by pollution~\cite{mahajan_citizen_2020, hsu2017community}. A significant amount of research has shown that the main producers of urban pollution are highly concentrated in disadvantaged, poor, and minority communities~\cite{Lubell2006}. Therefore, these communities have the most potential to drive fruitful policy development if they are engaged in the process effectively.

Community engagement regarding air pollution has historically been a challenge due to the high cost of air monitoring technology~\cite{commodore2017community}. However, the emergence of low-cost sensors creates opportunities for community-based organizations to collect data and perform community science~\cite{commodore2017community}. Data by itself lacks the power to inform communities, and thus data communication becomes a foundational challenge in driving engagement. Interactive data visualization provides a powerful outlet to present this data to these communities, empowering those affected to advocate for themselves. Visualization becomes a tool that can educate the people, allowing them to inform each other, which ultimately helps inform decision-making. This develops a combination of knowledge, skills, values, and motivation to make a difference and allows for robust civic engagement and participation.

CleanAirNowKC (CAN-KC) is an environmental justice organization based in Kansas City (KC), Kansas. Their goal is to work with communities threatened by air pollution to increase their participation in policymaking decisions and improve the health and quality of life in those communities. To date, CAN-KC has deployed nine PurpleAir air quality sensors in different locations throughout the community. The sites were chosen based on locations the community participants deemed most concerning, with plans to add more. Creating an engaging visualization that is both effective and uncluttered is a non-trivial task, which is compounded by real-time data processing. This paper details the implementation of an interactive map that will soon be deployed to the CAN-KC website. To build the visualization, a combination of web technologies (namely D3.js and Leaflet.js) are used in conjunction with the PurpleAir API for data collection. We aim to design an interactive map to visualize air quality sensor data that can facilitate the following tasks:
\begin{enumerate}[noitemsep,itemsep=4pt]
    \item Display real-time air quality data in these communities;
    \item Convey the impact of poor air quality on locals’ health;
    \item Visualizes the unequal burden of air pollution occurring in Kansas City fence line communities; and
    \item Provide visualizations that encourage community engagement, civic participation, community-led solutions in policymaking, and accountability of polluting industries.
\end{enumerate}

This work seeks to create a resource that will assist CAN-KC in organizing and reporting data and engage Kansas City decision-makers. By creating this resource, we hope that we can contribute to the efforts being made by CAN-KC to improve the air quality policy of Kansas City and perhaps inspire other municipalities to take similar action.

\section{Background}
\subsection{Air Monitoring Technology}
There have been numerous significant works on air monitoring technology in recent years. 
Baldauf et al.~\cite{baldauf2008traffic} presented a study on highly time-resolved characterization of traffic activity, meteorology, and air quality at varying distances from the road. They used several air quality parameters that represent the complex mixture of pollutants emitted by motor vehicles. Jiao et al.~\cite{jiao2015field} developed an integrated air and weather measurement system that was more easily deployed in outdoor community spaces, minimizes operator maintenance, and provides real-time, quality-checked data to the public. A system called MAAV was developed by Moore et al.~\cite{moore2018managing} to perform three tasks---measure air quality, annotate data streams and visualize real-time PM2.5\footnote{PM2.5 is a pollution measurement of the density of particulate with a diameter of 2.5 micrometers or smaller.} levels. The system used indoor and outdoor air quality monitors to capture PM2.5 levels. To inspect the measured data, they built a tablet-based interactive visualization.

Historically, air quality sensors were expensive. Technological advancement has led to affordable air quality sensors, empowering individuals and community organizations to monitor their local air quality. In this work, we utilize the low-cost PurpleAir sensor~\cite{feenstra2019performance}. PurpleAir uses laser particle counters to measure air pollutants and transmit the data via WiFi to PurpleAir servers.

\subsection{Air Quality Visualization Systems}

Several prior visualization tools have inspired our design. 

In 2017, Lu et al.~\cite{lu2017interactive} introduced an interactive web map application, built using web.py, Leaflet.js, and DS.js, that focused on the air quality of the leading cities of China. The data were retrieved from the national real-time air quality reporting system, which was reported hourly. Their observations detected several locations in China that had alarming air quality throughout the year.

Chen~\cite{Chen2019} used Google Earth and Keyhole Markup Language (KML) functionality to produce a heatmap visualization based on air quality readings. The heatmap ranged from blue (low concentration of pollutants) to red (high concentration of pollutants). Users could click on points of interest, which triggered a popup that displayed a 24-hour histogram of air quality, along with current conditions and considerations that should be taken under those conditions.

Cleary et al.~\cite{cleary2017making} created a web-based visualization, called Weave, that personalized air quality by relating the data to users' health, specifically targeting health hazards caused by the air pollution-related to traffic for members of the Boston Chinatown community. The tools used previously collected data (wind speed, wind direction, temperature, and traffic volume) and a statistical prediction model with interactively adjustable parameters to visualize how traffic influenced PM2.5 concentrations.

Finally, Nurgazy et al.~\cite{Nurgazy2019} extended the idea of data personalization by developing user profiles that informed context-aware visualizations. By using air quality data provided by the city of Melbourne, Australia, they developed an air pollution map that would adjust the visualization based on the user's reported sensitivities, included user age, color vision impairments, and sensitivity to pollutants. The website of a Swiss air quality technology company IQAir~\cite{iqair} includes a map-based visualization of air quality of different countries and regions of the world. Although it is capable of conveying air quality related information to anyone who is interested, a system like this does not provide a community level focus, therefore not necessarily meeting the goals of any  specific community.

\subsection{Community Empowered Air Quality Monitoring}

There have been several initiatives around the world to engage communities in improving air quality. Carvlin et al.~\cite{carvlin2017development} developed a community-engaged research study to provide real-time particulate matter (PM) air quality information at a high spatial resolution in Imperial County, California. The air quality network provided information on susceptible populations, assisted in identifying air pollution hotspots, and increased community awareness of air pollution. Hsu et al.~\cite{hsu2017community} collaborated with Allegheny County CleanAirNow community to develop an air monitoring system. Their focus was to help the community gather enough scientific evidence to support policymaking. The system built in this study was able to generate animated smoke images, air quality data, and crowd-sourced smell reports.  The result of their survey suggested that the system provided scientific evidence that could work in favor of the community in policymaking. However, the evaluation was done based on a small number of participants, all well-educated.

Other approaches have tried engagement techniques through social networks and gaming to improve air quality awareness.
Niemeyer et al.~\cite{niemeyer2009black} developed a collective networked public air pollution sensor for use within a game context for high school students in Los Angeles. The main goal of the game is to transform students into agents of change by making real-world measurements of air quality, tracking down the sources of pollutions, and exploring their impact on the environment around them.  Kim and Paulos~\cite{kim2010inair} developed a tool, ``inAir,'' for measuring and having visualizations of indoor air quality within a social network.  The study showed how simple visualizations of information undetectable to humans, such as indoor particle counts, can play a significant role in increasing people's awareness and understanding of air quality.

Finally, Hooker et al.~\cite{hooker2007pollution} developed an electronic street sign, called a ``Pollution e-Sign,'' using a form of hacking called bluejacking, which exploits weaknesses in Bluetooth. The system communicated local air quality information to passing devices without their permission to improve awareness.

\subsection{Kansas City Air Pollution}
Kansas City includes low-income communities of color fenceline to industrial sources of pollution, including the second-largest freight-car classification rail yard in the United States~\cite{atlantic}. With very limited accountability for environmental enforcement, Kansas City faces cumulative exposure to air pollutants.
 
CAN-KC and community members from Argentine, Kansas (a neighborhood in Kansas City) collectively gathered air pollution data using low-cost sensors in 2015-2016 to address community air pollution concerns, specifically regarding elemental carbon emitted by diesel engines used in the rail yard~\cite{cankc2015}. One-third of the air monitors measuring elemental carbon in November 2013 were reported to be above a level associated with short-term health risks. The community’s concern was enough to spark a one-year EPA study called KCTRAQS~\cite{kimbrough2019kansas}. The analyses conducted by the EPA are still being released. In partnership with CAN-KC, local academics are using the EPA KC-Traq air quality data to assess for a correlation between local air quality and health.

This particular project began in 2020, when members of the Kansas City-based environmental justice organization, CAN-KC, were connected to researchers at the University of South Florida through the Union of Concerned Scientists.

\section{Methods}

Our overall design goal was to provide Kansas City community members with access to current air quality information~(D1), as well as information about the health impact of bad air quality~(D2). Additionally, we provide access to recent historical~(D3) air quality data, including possible sources of pollution, to help to draw the attention of the policymakers to take necessary steps for improvement. In addition, we wanted to give community members the opportunity to actively participate in the process by reporting first-hand pollution information~(D4), e.g., industrial emissions, toxic releases, etc. Finally, we wanted the interactive map to enable CAN-KC members to gather evidence of pollution to guide decision-making~(D5).

\subsection{Air Quality Data Collection}
The data to be visualized is reported by numerous PurpleAir air quality sensors that the CAN-KC, in partnership with community members, placed around Kansas City. These sensors are connected via WiFi, and the sensor readings are made available through the PurpleAir API. The request frequency limit is relatively low~(max $1\ req./min.$). To properly provide pseudo-live readings, front-end requests are routed through a python-based proxy server, which refreshes the data from PurpleAir every 10 minutes. For development and evaluation, we captured a static dataset consisting of readings from 8 live PurpleAir sensors over 2 days.

\begin{figure}[!t]
    \centering
    \includegraphics[trim=0 40pt 0 26pt, clip, width=0.95\linewidth]{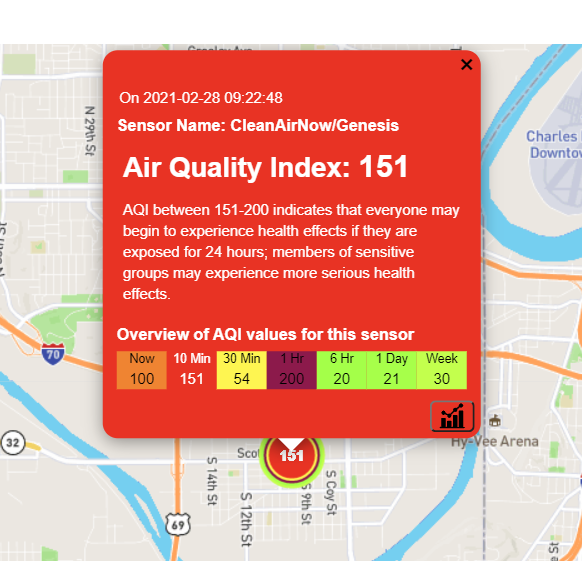}
    \caption{The info card is displayed when a sensor is clicked on.}
    \label{fig:info_card}
\end{figure}

\subsection{Air Quality Map}
The main feature of this data visualization is the map~(see \autoref{fig:map}). Using Leaflet.js, we create the map with MapBox tiles and OpenStreetMaps data as the primary users are local residents evaluating their surroundings in Kansas City. Therefore, a street-level map is the most helpful layer set. To summarize the data available and provide a central location for map interaction and interpretation, we have included a small overlay in the bottom left corner of the map. This overlay offers several functionalities. First, it allows the user to select which sensor data is displayed on the map~(D1). Additionally, a user may choose a specific sensor by name, which will zoom the map to the location of that sensor. A ``Find Me!'' button focuses the map on the user's location. Finally, it provides the ability to submit a report of transient air quality and other pollution events~(D4).

\subsection{Air Quality Glyphs}
To provide current air quality information~(D1), the most recent data for each sensor is represented with a circle marker at the reported sensor latitude and longitude. Additionally, the most recent PM2.5 reading is used to fill the marker using a linear color scale~(using a stoplight metaphor---green to yellow to red) shown in the lower right corner of the map display. A numeric representation is also appended to the center. This approach allows for both a cursory visual inspection of the map and specific real-time reading. For recent historical data~(D3), different colored rings are placed around each marker that indicates the PM2.5 reading over different intervals. With the map fully rendered, users can zoom in and pan to explore the other parts of the city.

\subsection{Sensor Popups}
\subsubsection{Info Card}
An information popup~(see \autoref{fig:info_card}) appears when the sensor marker is clicked. This information card gives the user an overall summary of the selected sensor. It shows basic information, including date, the value of the chosen metric~(PM2.5\_10 minute by default), what the Air Quality Index~(AQI)\footnote{Air Quality Index is a nationally uniform index for reporting air quality developed and promoted by the US Environmental Protection Agency (EPA).} value means for health and safety concerns~(D2), and the sensor name. It also shows the average AQI value over different time intervals~(D3).

\begin{figure}[!b]
  \centering
  \includegraphics[trim = 0 12pt 0 5pt, clip, width=0.95\linewidth]{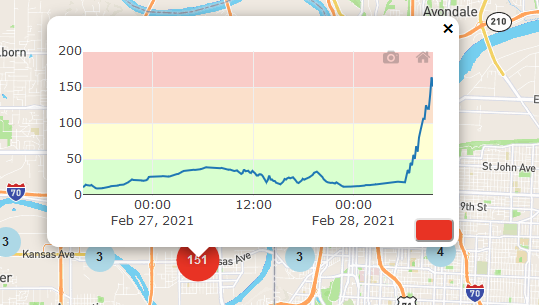}
  \caption{The secondary popup line chart view.}
  \label{fig:line_chart}
\end{figure}

\subsubsection{Line Chart}

A button in the bottom right corner of the sensor popup~(see \autoref{fig:info_card}) lets the user switch to a secondary historical line chart view~(D3). This view allows power users to access additional details. The line chart, generated using PlotlyJS, provides an overview of the sensor readings for the selected metric~(default PM2.5\_10Minutes) over time~(see \autoref{fig:line_chart}).  The background of the line chart is filled with the corresponding air quality color scale to enable visual correlation to the health risk level of that pollution~(D2).  The line chart is interactive, offering the ability to adjust the scales, zoom in with a bounding box, and tooltips that reveal specific air quality data corresponding to their mouse position. After adjusting the view, there is also an option to save the graph as an image, providing a method for gathering and sharing information for community engagement~(D5).

\subsection{Hazardous Waste Facilities}

\begin{figure}[!ht]
  \centering
  \begin{minipage}[b]{0.975\linewidth}
    \centering
    \includegraphics[trim= 0 70pt 0 0 , clip, width=0.9\linewidth]{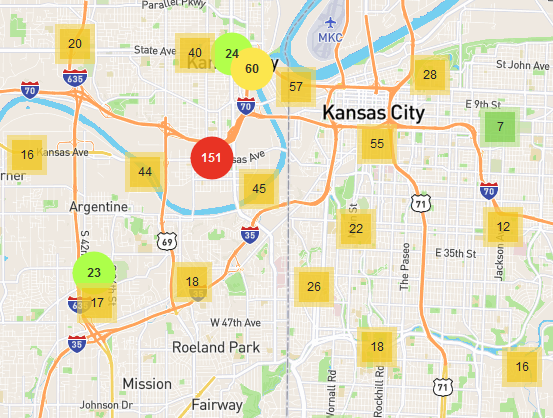}
  \end{minipage}
  \begin{minipage}[b]{0pt}
    \hspace{-95pt}
    \vspace{55pt}
    \fbox{\includegraphics[width=85pt]{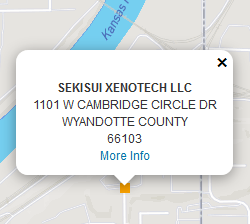}}
  \end{minipage}
  
  \caption{The map with hazardous waste layer and the popup displayed for hazardous waste sites~(upper right).}
  \label{fig:hazards}
\end{figure}

Superfund and other hazardous waste sites are included as a third layer on the map~(off by default to reduce information overload, see \autoref{fig:hazards}), sourced from the EPA\footnote{https://enviro.epa.gov/facts/rcrainfo/search.html}. There are numerous hazardous waste sites in Kansas City; individual points are clustered into markers to designate a larger area. Upon zooming in, the clusters are dismissed in favor of the localized hazardous waste sites. Users may click on individual markers~(\autoref{fig:hazards} upper right) to see the name of the primary contact of the waste site and its address, as well as a link to the EPA entry on that location~(which details the type of waste emitted and local contacts regarding disposal). This information serves two purposes: first, it allows Kansas City residents to correlate their localized air quality with its potential contributors~(D1), and second, it guides CAN-KC organizers regarding where new sensors may be best placed~(D5).

\section{Evaluation}
The visualization has been designed as a community engagement tool. The main objective of this visualization is to stimulate community engagement and community-led solutions in policymaking. As the primary users of this visualization are community members and the CAN-KC organizers, we asked them to use the tool and subsequently complete a survey to evaluate the effectiveness of this tool. We received 20 responses from CAN-KC organizers and members. The ages of these community members range between 18-84 years, 9 female, 11 male. Of the survey participants, 1 was involved in the website development process and is a co-author of this paper.

The evaluation was designed to probe 3 main questions. The first section focused on identifying how participants accessed the map to understand if the tool works properly on different machines and browsers. The second section asked the users to perform specific tasks to gauge the intuitive usability of the map. The final section evaluates the overall user experience. The overall feedback from the participants was very positive. 

\paragraph{Access Method} Most participants used a computer to access the map, but a few people used mobile phones as well. Participants using a computer did not face any technical issues. However, on mobile devices, some user interaction issues were identified, e.g., overlapping components, components too large for the screen, etc. 

\paragraph{Task Completion} Most users were able to access the map and find their location on the map with ease and retrieve information regarding the air quality impact on health. However, 5 participants found it extremely difficult to locate themselves. 
This indicates that (D1) and (D2) were largely accomplished.
When the participants were asked about the different colored rings around the sensor markers, 12 participants were unclear what it indicates. This feedback suggests that (D3) was not fully satisfied, and we either need to redesign our glyph or provide an explanation of it on the map.

\begin{figure}[!b]
  \centering
  \begin{minipage}[b]{0.375\linewidth}
      {\includegraphics[trim=5pt 10pt 69pt 0, clip, angle=90,width=\linewidth]{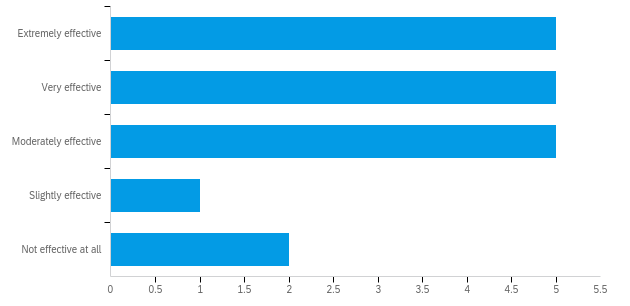}}
    \end{minipage}
    \hspace{15pt}
  \begin{minipage}[b]{0.425\linewidth}
    \centering
      {\includegraphics[trim=15pt 110pt 285pt 85pt, clip, width=\linewidth]{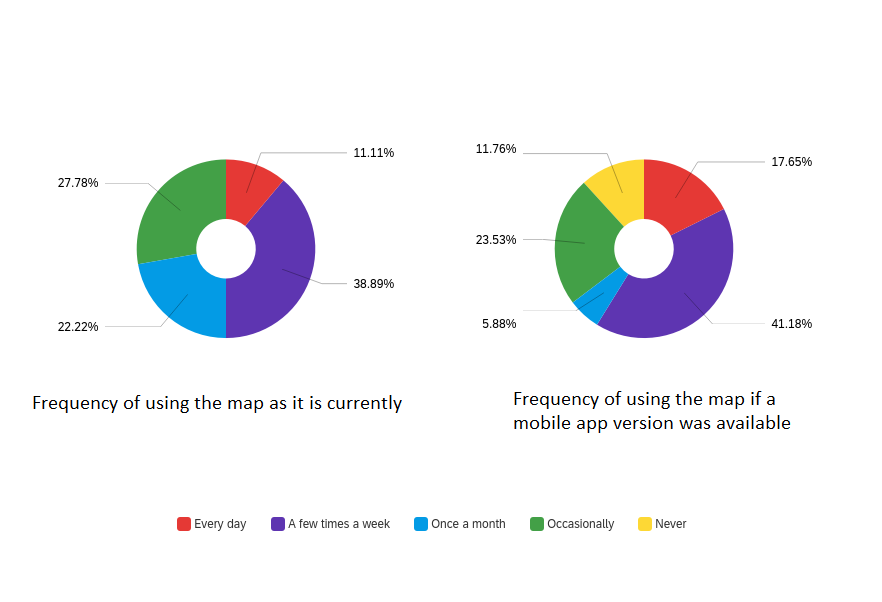}}
      {\includegraphics[trim=270pt 100pt 30pt 85pt, clip, width=\linewidth]{figures/donut_chart_usage.png}}
      {\includegraphics[trim=105pt 40pt 295pt 310pt, clip, height=7pt]{figures/donut_chart_usage.png}}
      
      \vspace{-3pt}
      {\includegraphics[trim=245pt 40pt 115pt 310pt, clip, height=7pt]{figures/donut_chart_usage.png}}
    \end{minipage}  
  \caption{Left: Feedback regarding the effectiveness of the map at communicating air pollution data in Kansas City. Right: Feedback regarding the frequency of using the map.}
  \label{fig:effective}
\end{figure}

\paragraph{Overall Experience} When the participants were asked if the map could effectively communicate air pollution data in Kansas City, most agreed. The overview of the feedback for this question is showed in \autoref{fig:effective} left. This result is an indirect indication that the system would be useful for (D5). Participants responded positively when asked if they would use this map to monitor the air quality. Since one important aspect of (D4) was that we wanted to increase community engagement and participation, the responses we received indicated that the system could attain that. We also asked the participants how often they would access the map if a mobile application were available to monitor the air quality. Although the response rate was similar to the desktop version, 2 participants reported they would never use the mobile app to check the map. The contrast of the responses regarding the frequency of usage is shown in \autoref{fig:effective} right. When they were asked if they believed that the Kansas City community would benefit from having this map available online, 16 participants strongly agreed and 1 participant somewhat agreed. In the survey, 2 community members mentioned that they had trouble understanding what different intervals of data collection mean, suggesting either an explanation should be provided or that this feature is too technical for a general audience.

\section{Conclusion \& Future Work}
In this paper, we detail an interactive map designed in partnership with the environmental justice non-profit CAN-KC. The goal of the map is two-fold: to interactively display local air pollution data and inform the community members about air pollution in their community. The map displays information regarding the current and historical air quality values read from PurpleAir sensors and provides an interpretation of what health actions should be taken. Additionally, users can monitor and identify industrial pollution sources. We will make the website live in the near future. Use of this map, in unison with educational events and attendance to city-directed meetings, can serve to improve community engagement and involvement in local policy. Once these resources are available to the general public, we will then be able to evaluate how successful our visualization is at informing and engaging the community. 

One challenge of working with the community to develop a tool was that most of them are unfamiliar with any technical aspect of the system. It made it challenging to elicit requirements, but involving CAN-KC organizers in the development process was critical to understanding the requirements before making the design decisions.

Based on the feedback we have received, we plan to update the CAN-KC website with better explanations and easier-to-understand and more user-friendly visualizations of data collection intervals, industrial pollution sources, and the meaning of AQI. We also plan on making the map more suitable for mobile devices and fix the existing user interface issues. We want to improve the functionality of pollution reporting for users. Having a mobile app version of this map, we also believe, would make submitting reports easier.

We believe this will increase community engagement and also help to collect evidence against major contributors to pollution that can play a significant role in policymaking.

\acknowledgments{This work was partially supported by NIH-5R34HL145442-02.}

\bibliographystyle{abbrv-doi}
\bibliography{main}

\end{document}